\documentclass[aps,prl,twocolumn,showpacs,groupedaddress]{revtex4}
\usepackage{graphicx}

\begin{document}


\title{Sagnac effect in a chain of mesoscopic quantum rings }

\author{Chris P. Search, John R. E. Toland, and Marko Zivkovic}
\affiliation{Department of Physics and Engineering Physics,Stevens Institute of Technology, Hoboken, NJ 07030}

\date{\today}

\begin{abstract}
The ability to interferometrically detect inertial rotations via the Sagnac effect has been a strong stimulus
for the development of atom interferometry because of the potential $10^{10}$ enhancement of the rotational phase shift in comparison
to optical Sagnac gyroscopes. Here we analyze ballistic transport of matter waves in a one dimensional chain of $N$ coherently coupled quantum rings
in the presence of a rotation of angular frequency $\Omega$. We show that the transmission probability, $T$, exhibits zero transmission stop gaps as a function of the rotation rate interspersed with regions of rapidly oscillating finite transmission. With increasing $N$, the transition from zero transmission to the oscillatory regime becomes an increasingly sharp function of $\Omega$ with a slope $\partial T/\partial \Omega\sim N^2$. The steepness of this slope dramatically enhances the response to rotations in comparison to conventional single ring interferometers such as the Mach-Zehnder and leads to a phase sensitivity well below the standard quantum limit.
\end{abstract}

 \maketitle

Matter wave interferometry is a uniquely quantum mechanical effect that has been a subject of great interest since the early days of quantum theory. Besides simply probing fundamental quantum phenomena such as complementarity, matter wave interferometers (MIs) and in particular atom interferometers (AIs), have been shown to have a number of important applications including inertial navigation, geophysical prospecting, and tests of general relativity \cite{Gustavson,McGuirk,Durfee}. One of the most important applications is the ability of a MI to detect inertial rotations via the Sagnac effect. For a MI using particles of mass $M$ and enclosing an area $A$, the phase shift between counterpropagating waves is $\Theta_{S}=2M\Omega A/\hbar$ where $\Omega$ is the rotation rate perpendicular to the plane of the interferometer \cite{Clauser}. For typical atomic masses, this is $Mc^{2}/\hbar \omega \sim 10^{10}$ larger than optical Sagnac based interferometers with light of frequency $\omega$ \cite{Chow,Culshaw}. Current generation laboratory AI's \cite{Gustavson,Durfee,McGuirk,lenef} already approach the sensitivity of the best commercially available optical Sagnac gyroscopes, which have in large part replaced mechanical gyroscopes for inertial navigation and positioning-stabilization applications \cite{Chow,Culshaw}.

One of the main disadvantages of current AI's is the small enclosed area, which effectively limits the rotation induced interferometric fringe shift. For example, in a recent experiment \cite{Durfee} the enclosed area of the gyroscope is only $A\approx 22mm^2$ despite having an overall length of $\sim 2m$. By contrast, a typical optical fiber gyroscope has an effective area $A\sim100m^2$ in a much more compact form factor \cite{Culshaw}. Further progress for AI's would require a configuration where the sensitivity can be enhanced by creating a larger effective area while maintaining compact size interferometers.

\begin{figure}[htb]
\centering
\includegraphics[height=8cm, width=0.4\textwidth]{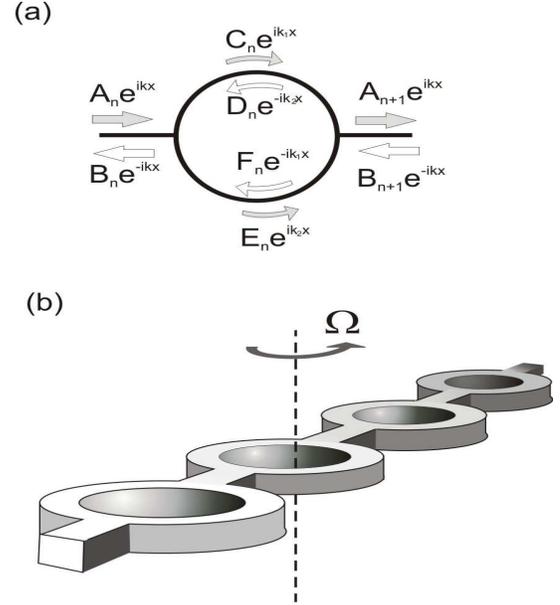}
\caption{(a) Schematic diagram of a ballistic chain of $N$ ring shaped matter wave interferometers.
(b) Diagram of a single ring of radius $r_0$ in the chain connected via $Y$ beam splitters to
1D ballistic wave guides. Labeled are the wave functions for the upper and lower arms as well as
the incident and reflected wave functions at the input and output wave guides.
}\label{PIC.1}
\end{figure}

Here we study transport of ultra-cold atomic matter waves through a one dimensional chain of $N$ rings in the presence of an inertial rotation of frequency $\Omega$, as depicted in Fig. 1. Each of the rings behaves like a miniature Sagnac interferometer with a single input and output port created by  $Y$-beam splitters. As a result of destructive interference betweeen rings, the transmission through the entire chain exhibits periodic transmission gaps of zero transmission for specific $\Omega$. At the edges of the gaps, the transmission as a function $\Omega$ becomes step like for large $N$ and jumps from zero to unit transmission and subsequently exhibits increasingly rapid oscillations for increasing $N$ inside of the transmission window. We will show that such a structure can be used to construct an exceptionally sensitive matter wave Sagnac interferometer since the slope of the transmission at the edges of the transmission gaps is large and proportional to $N^2$. We base our analysis on that of non-interacting particles using the transfer-matrix approach commonly used in mesoscopic transport \cite{Davies}.

Such a matter wave interferometer chain could be readily implemented using so-called atom chips \cite{fortagh}. Atom chips use currents through lithographically defined wires on standard semiconductor substrates to create magnetic waveguides above the chip for the trapping and guiding of ultra-cold atoms.  Such magnetic atom chips have already been used to demonstrate a $Y$ beam splitter \cite{Cassettari}, which is the essential element of our device. There have also been demonstrations of beam splitters and interferometers with optical atom chips that use micro-lenses to create configurable optical dipole potentials above the chip surface \cite{Dumke}. Alternatively, one could conceive of an optical lattice made of toroidal optical dark ring traps, as recently demonstrated \cite{olson,courtade}.

Finally we note that the optical Sagnac effect in ring shaped arrays of coupled resonator optical wave guides (CROWs), have been analyzed as potential micro-optical gyroscopes \cite{Scheuer}. They also predict an enhancement to the Sagnac effect that depends on the number of CROWs in the array. However, these devices are impractical because of the small dimensions so that the rotation rates studied were in the range $\Omega\sim10^2-10^8 s^{-1}$ while for navigation, one must be able to measure $\Omega\sim 5 \times 10^{-8}s^{-1}$ \cite{Culshaw}.

A sketch of the proposed device is shown in Fig. 1.  We assume that all of the rings and interconnecting segments in the chain are identical so that the structure
can be viewed as a periodic lattice of interferometers. Figure 1(a) shows the relevant wave functions for a single ring segment. We assume that the transverse confining potentials that form the waveguides are large enough that particles only propagate in the lowest transverse mode.

The transfer matrix approach allows us to construct the total transmission probability for the chain as the product of transmission matrices for each individual segment. The wave functions for the input of the $n^{th}$ ring, the upper and lower arms, and the output (input) of the $n^{th}$($(n+1)^{st}$) are $\psi_n=A_ne^{ikx}+B_ne^{-ikx}$, $\psi_n^{(L)}=E_ne^{ik_2x}+F_ne^{-ik_1x}=(E_ne^{ikx}+F_ne^{-ikx})e^{-i2m A\Omega x /\hbar L}$, $\psi_n^{(U)}=C_ne^{ik_1x}+D_ne^{-ik_2x}=(C_ne^{ikx}+D_ne^{-ikx})e^{i2mA\Omega x/\hbar L}$, $\psi_{n+1}=A_{n+1}e^{ikx}+B_{n+1}e^{-ikx}$. Here we have incorporated the Sagnac phase shift into the wave numbers for the ring $k_{1,2}=k\pm2mA\Omega/\hbar L$, which can be obtained either from the mathematical equivalence of the Sagnac and the Aharonov-Bohm effect \cite{sakurai} or by consideration of the energy shifts of the clockwise and counterclockwise modes induced by the rotation \cite{Chow}. Here $L$ is the circumference of the ring of radius $r_0$ and area $A$. For a single ring, the transfer matrix is obtained by matching the wave functions and their derivatives (current conservation) at the Y-splitters \cite{xia}, so that after elimination of $C_n$, $D_n$, $E_n$, and $F_n$ the transfer matrix relating the $n^{th}$ and $(n+1)^{st}$ segments, $(A_{n+1},B_{n+1})^T={\mathcal T}_{n+1,n}(A_n,B_n)^T$ has the form \cite{cui}:
\begin{equation}
\mathcal{T}_{n+1,n}=\frac{1}{2\cos\Theta_S \sin(kL)}
\left(
\begin{array}{cc}
\alpha & \beta \\
\alpha^* & \beta^* \\
\end{array}
\right)
 \left(
\begin{array}{cc}
e^{ikd} & e^{-ikd} \\
-ie^{ikd} & ie^{-ikd}
\end{array}
\right)
\end{equation}
where $\alpha=\cos(kL)\sin(kL)+2i(\cos^2(\Theta_S)-\cos^2(kL))$ and $\beta=-\sin^2(kL)/2+i\cos(kL)\sin(kL)$. $d$ is the length of the segments
connecting the rings.

The total transmission probability is given by $T=|t|^2=1/|T_{11}|^2=1/|T_{22}|^2$ where $T_{ij}$ are the elements of the total transmission matrix $\mathcal{T}(N)=\prod_{n=1}^{N}\mathcal{T}_{n+1,n}=\mathcal{T}^N$.
The product of $N$  $2\times 2$ matrices can be expressed as $\mathcal{T}^N=u_{N-1}(\epsilon)\mathcal{T}-u_{N-2}(\epsilon)I$
where $I$ is the identity matrix, $\epsilon=\frac{1}{2}Tr[\mathcal{T}]$, and $u_N$ is the $N^{th}$ order Chebyshev polynomial of the second kind. The transmission probability is then
\begin{equation}
T=\frac{1}{1+C(kL,\Theta_S)u^2_{N-1}(\epsilon)} \label{transmission}
\end{equation}
where
\begin{equation}
C(kL,\Theta_S)=\frac{[ 4\sin^2(\Theta_S) -3\sin^2(kL)]}{16\sin^2(kL)\cos^2(\Theta_S)}
\end{equation}
and $\epsilon=\cos(kd)\cos(kL)/\cos(\Theta_S)+\sin(kd)(4\sin^2(\Theta_S)-5\sin^2(kL))/(4\cos(\Theta_S)\sin(kL))$.


We consider a stream of $n$ uncorrelated atoms incident on the interferometer. This implies that the incident atomic flux, $J$,
is small enough that quantum statistics and mean field interactions can be ignored. For a 1D beam, this results in the requirement that $\lambda^2<h/(mJ)$ where $\lambda=2\pi/k$ is the atomic de Broglie wavelength \cite{search-noise-limits}. In this case the transmission statistics through the interferometer is given by a binomial distribution with the average number of transmitted and reflected (backscattered) particles are defined as $\langle \hat{n}_T\rangle=nT$ and $\langle \hat{n}_R\rangle=n(1-T)$, respectively. The variances are then $\langle \Delta\hat{n}_T^2\rangle=\langle \Delta\hat{n}_R^2\rangle=nT(1-T)$.

The shot noise in the transmitted current through the chain that results from the stochastic partitioning of incident particles into transmitted and reflected beams represents the fundamental limit to the phase sensitivity. The phase sensitivity is given by $\Delta\Theta_S=\sqrt{\langle \Delta \hat{n}_T^2\rangle}/|\partial \langle \hat{n}_T\rangle/\partial \Theta_S|$ \cite{search-noise-limits}, which in our case yields
\begin{equation}
\Delta\Theta_S=\frac{1}{\sqrt{n}}\frac{\sqrt{T(1-T)}}{|\partial T/\partial \Theta_S|} \label{phase-sensitivity}
\end{equation}
where the first term $\sim 1/\sqrt{n}$ is the familiar term representing the standard quantum limit (SQL) for uncorrelated particles \cite{giovannetti}. This expression is readily interpreted as the minimum detectable change in the phase $\Delta\Theta_S$ for which the change in the output signal $(\partial \langle \hat{n}_T\rangle/\partial \Theta_S)\Delta\Theta_S$ just equals the noise in the output, $\sqrt{\langle \Delta\hat{n}_T^2\rangle}$.

\begin{figure}[htb]
\centering
\includegraphics[height=7cm, width=0.5\textwidth]{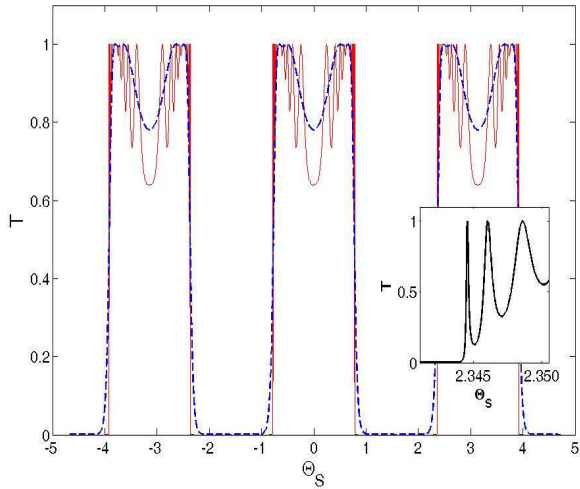}
\caption{(Color Online) Transmission probability, $T$, through a chain of $N$ rings for $kL=0.25\pi$ and $kd=0$ as a function of the Sagnac rotational phase shift, $\Theta_S$. The red solid line is $N=50$ while the blue dashed line is $N=5$. The inset shows a close up of the behavior at the edge of one of the transmission windows for $N=50$.
}\label{PIC.1}
\end{figure}

The relevant quantity besides just the transmission is therefore $\partial T(\Theta_S)/\partial\Theta_S$. We will focus here on the case of rings that are directly coupled without any interconnects, $kd=0$. Figure 2 shows $T(\Theta_S)$ plotted for several values of $N$ and $kL=\pi/4$. One can see that the transmission is divided into periodic regions of large oscillating transmission close to 1 and those of zero transmission. The transition from zero to maximal transmission occurs when $|\epsilon|=|\cos(kL)/\cos(\Theta_S)|=1$ while the slope of the transmission at the edge of the window becomes larger and larger until in the limit of $N\rightarrow \infty$, T becomes a step function at $|\epsilon|=1$. The zero transmission stop gaps are due to the phase $\Theta_S$ acquired in each ring that leads to destructive interference between segments.

We can estimate $\partial T(\Theta_S)/\partial\Theta_S$ at the edge of the transmission window for large $N$ by noting that the behavior at the edge of the window is dictated entirely by the Chebyshev polynomial in Eq. \ref{transmission} rather than $C(kL,\Theta_S)$ since $C(kL,\Theta_S)$ oscillates much more slowly than $u_{N-1}$. $u_{N-1}$ has $N-1$ zeros in the transmission window leading to transmission resonances at $\epsilon=\cos(m\pi/N)$ for $m=1,...,N-1$ \cite{cui}.  $C(kL,\Theta_S)\rightarrow \infty$ when $\Theta_S=\pm \pi/2, \pm 3\pi/2,...$ but provided $kL\neq \pi/2, \pm 3\pi/2,...$, this occurs inside of the transmission stop gap created by $u_{N-1}^2$ and therefore does not further contribute to the inhibition of transmission. In the opposite extreme, the locations where $C(kL,\Theta_S)=0$ occur when $\sin^2(\Theta_S)/\sin^2(kL)=3/4$, which can be shown to only occur inside the transmission window and never at the edge. Therefore it is safe to take $C(kL,\Theta_S)\neq 0,\infty$ when $|\epsilon|\approx 1$. Due to the periodicity of $T$ we can focus on the interval $0<\Theta_S<\pi/2$ where the transmission jumps from $T(\Theta_{S,1})=1 \rightarrow T(\Theta_{S,0})=1/(1+C(kL,\Theta_{S,0})N^2)$ where $\cos(\Theta_{S,1})=\cos(kL)/\cos(\pi/N)$ is the zero of $u_{N-1}$ closest to $\epsilon=1$ while $\cos(\Theta_{S,0})=\cos(kL)$ corresponds to the beginning of the zero transmission stop gap at $\epsilon=1$. For $N\gg 1$, $T(\Theta_{S,0})=1/(1+C(kL,\Theta_{S,0})N^2)\approx 0$. The derivative is then $\partial T/\partial \Theta_S\sim (T(\Theta_{S,1})-T(\Theta_{S,0})/(\Theta_{S,1}-\Theta_{S,0})=1/(\Theta_{S,1}-\Theta_{S,0})$. $\Theta_{S,1}-\Theta_{S,0}=\cot(kL)(1-1/\cos(\pi/N))$, which for large $N\gg 1$ yields
\begin{equation}
\partial T/\partial\Theta_S\sim-(2/\pi^2)\tan(kL)N^2. \label{transmission-slope}
\end{equation}
Figure 3 shows the dependence on $N$ and $kL$ of the maximum value of $|\partial T/\partial\Theta_S|$ at the edge of the transmission window, which was extracted numerically from Eq. \ref{transmission}. One can see in Fig. 3(a) that for $N\gg 1$ the slope does indeed scale like $N^2$ while Fig 3(b) the dependence of $|\partial T/\partial\Theta_S|$ on $kL$ for fixed $N$ indicates a $\tan(kL)$ dependence. In both Figs. 3(a) and 3(b), the numerical data is fitted to $|\partial T/\partial\Theta_S|=\beta \tan(kL)N^2$, where $\beta$ is approximately a factor of 10 larger than $2/\pi^2$. The reason is that Eq. \ref{transmission-slope} measures the slope of the line that bisects the beginning the transmission stop gap with the first transmission maximum. Since $\partial^2 T/\partial\Theta_S^2>0$ this is smaller than the maximum of $|\partial T/\partial\Theta_S|$ between those points.

\begin{figure}[htb]
\centering
\includegraphics[height=10cm, width=0.5\textwidth]{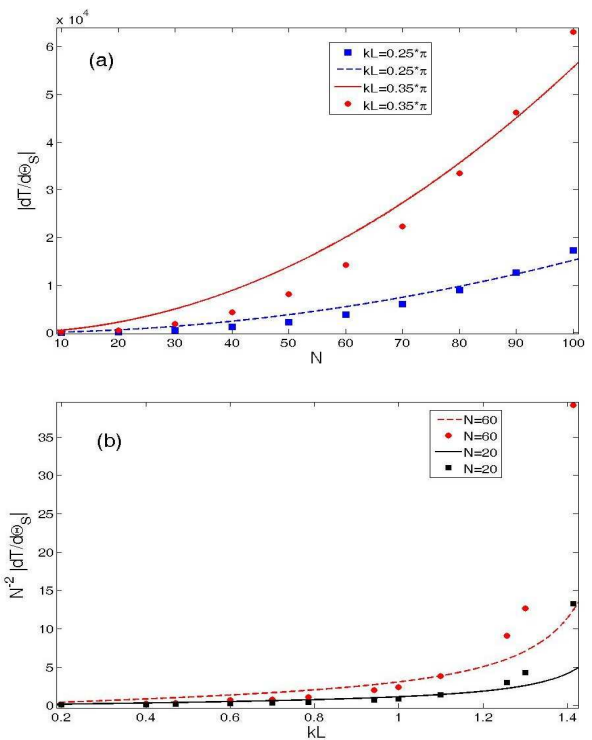}
\caption{(Color Online) Maximum slope of the transmission, $|\partial T/\partial \Theta_S|$, as a function of (a) the number of rings $N$ and (b) the total path length of the interferometer, $kL$. In both figures, circles and squares show data points obtained directly from Eq. \ref{transmission} while the lines show a fit to the function $|\partial T/\partial \Theta_S|=\beta \tan(kL)N^2$.
}\label{PIC.1}
\end{figure}

We estimate the phase sensitivity using $T(1-T)=C(kL,\Theta_S)u_{N-1}^2(\epsilon)/(1+C(kL,\Theta_S)u_{N-1}^2(\epsilon))^2\approx 1/CN^2$ for $|\epsilon|\approx 1$ and $N\gg 1$. Along with Eq. \ref{phase-sensitivity} and \ref{transmission-slope},  we find that
\[
\Delta\Theta_S\sim \frac{1}{\sqrt{n}N^3}.
\]
In a standard Mach-Zehnder interferometer, $\sqrt{\langle \Delta\hat{n}_T^2\rangle}$ is proportional to $\partial \langle \hat{n}_T\rangle/\partial \Theta_S$ resulting in a phase uncertainty equal to $1/\sqrt{n}$ that is independent of $\Theta_S$ and is typical of the SQL for an interferometer \cite{giovannetti}. This is no longer the case in our device since at the transmission window edges, $|\partial \langle \hat{n}_T\rangle/\partial \Theta_S|$ approaches infinity as $N\rightarrow \infty$ while at the same time the noise is limited to $T(1-T)\leq 1/4$. Even for modest values of $N$ the phase sensitivity can significantly surpass the SQL. Figure 4 shows the minimum value of $\sqrt{n}\times \Delta\Theta_S$ calculated numerically from Eq. \ref{transmission} at the edges of the transmission windows and fitted to the curve $\alpha N^{-3}$. One can see from the figure that for several tens of rings $\Delta\Theta_S\sim 10^{-5}n^{-1/2}$. The inset of Fig. 4 shows that $\sqrt{n}\times \Delta\Theta_S$ has minimums to either side of a maximum, which occurs at the same location as the transmission resonance since both $T(1-T)$ and $\partial T/\partial \Theta_S$ vanish at the peak of the resonance.

\begin{figure}[htb]
\centering
\includegraphics[height=6.5cm, width=0.5\textwidth]{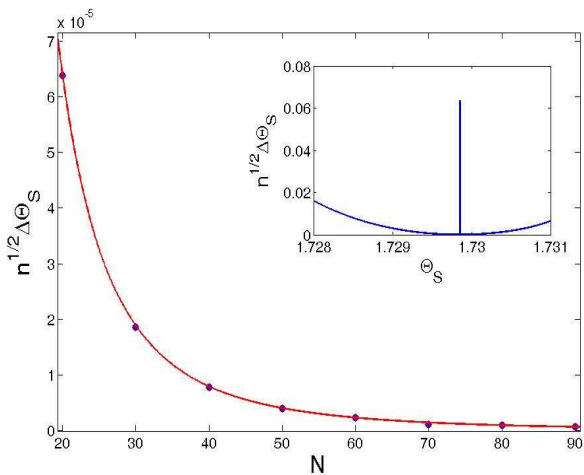}
\caption{(Color Online) The minimum value of $\sqrt{n}\times \Delta\Theta_S$, which represents the phase sensitivity relative to the standard quantum limit, $n^{-1/2}$, as a function of the number of rings, $N$ for $kL=0.45\pi$ and $kd=0$. Circles represent points calculated directly from Eq. \ref{transmission} while the red line is a curve fit $\alpha N^{-3}$. Inset shows $\sqrt{n}\times\Delta\Theta_S$ calculated from Eq. \ref{transmission} for $N=20$ rings as a function of $\Theta_S$.
}\label{PIC.1}
\end{figure}

To better relate these results to the detection of rotations, let us estimate how large the range $\Theta_{S,1}-\Theta_{S,0}$ is in terms of the range of rotation rates $\delta\Omega$ for atoms, $\delta\Omega=\hbar\pi\cot(kL)/4MAN^2$. Letting $\cot(kL)=1$ and considering $^{87}$Rb atoms in a rings of radius $10\mu m$ we obtain, $\delta\Omega=5.74N^{-2}s^{-1}$ for $N\gg 1$. This indicates that even for a modest number of rings, the device would be sensitive to sub-Hertz changes in $\Omega$. By contrast, a full period of $T(\Theta_S)$, which corresponds to a $\pi$ interval for $\Theta_S$, is equivalent to a rotation range of $\Delta\Omega=3.7s^{-1}$ independent of $N$.

Finally, we address the point that in reality one wants to measure changes in rotations relative to the inertial state $\Omega=0$ rather than changes in rotations relative to a finite rotation corresponding to the edge of a transmission window. Letting $kd\neq 0$ adds additional resonances but does not shift the stop gaps towards $\Omega=0$. However, by adding a slight asymmetry to the lengths of the upper and lower arms of each ring, $\Delta L$, the Sagnac phase shift in the transmission function is replaced with $\Theta_S\rightarrow \Theta_S+k\Delta L$. $\Delta L$ can be tuned to control the location of the transmission stop gaps.

In conclusion, we have studied the Sagnac effect in a coherent chain of matter wave interferometers and shown that the existence of transmission stop gaps could be used to enhance the phase sensitivity beyond the standard quantum limit. The universality of our transfer matrix approach implies that these results are also applicable to studying the Sagnac effect in ballistic conductors such as graphene or two dimensional electron gases \cite{Zivkovic}.
With some modification to the design, the same method could be used for matter wave gravity gradiometers.

\end{document}